\pdfoutput=1

\documentclass[sigplan]{acmart} 
\AtBeginDocument{%
  \providecommand\BibTeX{{%
    \normalfont B\kern-0.5em{\scshape i\kern-0.25em b}\kern-0.8em\TeX}}}

\usepackage{enumitem}

\usepackage{tikz,amsmath,amsthm,graphicx}
\usepackage[edges]{forest}

\usepackage{url}

\newcommand{\blue}[1]{\textcolor{blue}{{#1}}}

\newcommand{\eg}{\textit{e.g.}}

\newcommand{\ie}{\textit{i.e.}}
\newcommand{\et}{{ et al. }}

\setcopyright{none}
\begin{document}
\settopmatter{printacmref=false}
\renewcommand\footnotetextcopyrightpermission[1]{}
\pagestyle{plain}

\title{Combating Misinformation in the Age of LLMs: Opportunities and Challenges}

\author{Canyu Chen}
\affiliation{
  \institution{Illinois Institute of Technology}
  \city{Chicago IL}
  \country{USA}
}
  \email{cchen151@hawk.iit.edu}

\author{Kai Shu}
\affiliation{
  \institution{Illinois Institute of Technology}
  \city{Chicago IL}
  \country{USA}
}
\email{kshu@iit.edu}

\begin{abstract}
Misinformation such as fake news and rumors is a serious threat to information ecosystems and public trust. The emergence of Large Language Models (LLMs) has great potential to reshape the landscape of combating misinformation. Generally, LLMs can be a double-edged sword in the fight. On the one hand, LLMs bring promising opportunities for combating misinformation due to their profound world knowledge and strong reasoning abilities. Thus, one emergent question is: \textit{can we utilize LLMs to combat misinformation?} On the other hand, the critical challenge is that LLMs can be easily leveraged to generate deceptive misinformation at scale. Then, another important question is: \textit{how to combat LLM-generated misinformation?} In this paper, we first systematically review the history of combating misinformation before the advent of LLMs. Then we illustrate the current efforts and present an outlook for these two fundamental questions respectively. The goal of this survey paper is to facilitate the progress of utilizing LLMs for fighting misinformation and call for interdisciplinary efforts from different stakeholders for combating LLM-generated misinformation\footnote{More resources on ``\textbf{LLMs Meet Misinformation}'' are on the website: \blue{\url{https://llm-misinformation.github.io/}}}.
\end{abstract}
\maketitle


\section{Introduction}
Misinformation has been a longstanding and serious concern in the contemporary digital age~\cite{shu2017fake}. With the proliferation of social media platforms and online news outlets, the barriers to generating and sharing content have significantly diminished, which also expedites the production and dissemination of various kinds of misinformation (\eg, fake news, rumors) and exaggerates its influence at scale~\cite{zhou2020survey,DBLP:journals/csur/ZubiagaABLP18,10.1145/3534678.3542615,wu2019misinformation,kumar2018false,meel2020fake,lazer2018science,scheufele2021misinformation}.  As the consequence of prevalent misinformation, the public's belief in truth and authenticity can be under threat. Thus, there is a pressing need to combat misinformation to safeguard information ecosystems and uphold public trust, especially in high-stakes fields such as healthcare~\cite{chen2022combating} and finance~\cite{rangapur2023investigating}. 

The advent of LLMs~\cite{zhao2023survey} (\eg, ChatGPT, GPT-4~\cite{bubeck2023sparks}) has started to make a transformative impact on the landscape of combating misinformation. In general, LLMs are \textbf{\textit{a double-edged sword}} in the fight against misinformation, indicating that LLMs have brought both emergent \textbf{\textit{opportunities}} and \textbf{\textit{challenges}}. \textbf{On the one hand}, the profound \textit{world knowledge} and strong \textit{reasoning abilities} of LLMs suggest their potential to revolutionize the conventional paradigms of misinformation \textbf{\textit{detection}}, \textbf{\textit{intervention}} and \textbf{\textit{attribution}}. In addition, LLMs can be augmented with external knowledge, tools, and multimodal information to further enhance their power and can even operate as autonomous agents~\cite{xi2023rise}. \textbf{On the other hand}, the capacities of LLMs to generate human-like content,  possibly containing hallucinated information, and follow humans' instructions~\cite{he2023large} indicate that LLMs can be easily utilized to generate misinformation in an \textit{unintentional} or \textit{intentional} way. More seriously, recent research~\cite{chen2023llmgenerated} has found that LLM-generated misinformation can be harder to detect for humans and detectors compared to human-written misinformation with the \textit{same semantics}, implying that the misinformation generated by LLMs can have \textbf{\textit{more deceptive}} styles and potentially cause \textit{more harm}.

In this paper, we first provide a comprehensive and systematic review of the history of combating misinformation before the rise of LLMs with a focus on the detection aspect in Section~\ref{History of Combating Misinformation}. Then we delve into both the opportunities and challenges of combating misinformation in the age of LLMs. As for the opportunities, we will illustrate ``\textbf{\textit{can we
utilize LLMs to combat misinformation?}}'' in Section~\ref{LLMs for Combating Misinformation}. We will present the motivation for adopting LLMs in the fight against misinformation, the current efforts on utilizing LLMs for combating misinformation, which are mainly around the detection aspect, and an outlook embracing the intervention and attribution aspects. As for the challenges, we will discuss ``\textbf{\textit{how to combat LLM-generated
misinformation?}}'' in Section~\ref{LLM-Generated Misinformation}. We will dive into the characterization, emergent threats, and countermeasures of misinformation generated by LLMs. Looking ahead, we also point out the potential real-world devastating risks of LLM-generated misinformation in the near future, which may not be exhibited yet, and the desired interdisciplinary measures. Through this survey paper, we aim to \textbf{facilitate the adoption of LLMs in combating misinformation} and call for \textbf{collective efforts from stakeholders in different backgrounds to fight misinformation generated by LLMs}.

\section{History of Combating Misinformation}
\label{History of Combating Misinformation}
In this section, we conduct a systematic and comprehensive review of the techniques for detecting online misinformation before the emergence of LLMs to provide an overview of the history of combating misinformation in terms of the efforts on detection. Generally, we propose to categorize the detection methods into seven classes based on real-world scenarios: capturing linguistic features, leveraging neural models, exploiting social context, incorporating external knowledge, enhancing generalization ability, minimizing supervision cost, and fusing multilingual and multimodality.

\subsection{Capturing Linguistic Features} 
\label{Capturing Linguistic Features} 
Numerous linguistic features have been studied for differentiating misinformation from true information and can be roughly categorized as \textit{stylistic} features, \textit{complexity} features and \textit{psychological} features~\cite{aich-etal-2022-demystifying,horne2017just}. As for \textit{stylistic} features, prior research has found that misleading tweets are usually longer, use a more limited vocabulary, and have more negative sentiment~\cite{antypas-etal-2021-covid,rubin-etal-2016-fake}. Also, studies have shown that fake news tends to favor informal, sensational, and affective language style since it aims to attract readers' attention for a short-term financial or political goal~\cite{przybyla2020capturing,allcott2017social,bakir2018fake}. It is discovered that misleading articles use more swear words, subjective terms, superlatives, and modal adverbs to exaggerate a piece of news~\cite{rashkin2017truth}. As for \textit{complexity} features, misinformation is likely to be linguistically less complex and more redundant~\cite{antypas-etal-2021-covid} when measured by textual lexical diversity (MTLD) and type-token ratio (TTR)~\cite{mccarthy2005assessment}. The typical \textit{psychological} features are based on word counts correlated with different psychological processes and basic sentiment analyses such as  Linguistic Inquiry and Word Count (LIWC)
dictionaries~\cite{tausczik2010psychological},  which are shown to be strongly associated with the possibility of being misleading~\cite{mahbub2022covid}. 
Based on the linguistic patterns, multiple detectors are proposed~\cite{levi-etal-2019-identifying,aich-etal-2022-demystifying,mahyoob2020linguistic,mahyoob2020linguistic,garg2022linguistic,azevedo-etal-2021-lux,karimi-tang-2019-learning,prezrosas2017automatic}. For example, Mahyoob\et proposed to leverage 16 linguistic attributes, which include lexical, grammatical and syntactic features, to identify the nuance between fake and factual news~\cite{mahyoob2020linguistic}.

\subsection{Leveraging Neural Models}
\label{Leveraging Neural Models}
With the development of deep learning in natural language processing, more recent works utilize neural models such as Long Short-Term Memory (LSTM)~\cite{hochreiter1997long} and Convolutional Neural Network (CNN)~\cite{kalchbrenner2014convolutional} for feature extraction and prediction instead of manually extracting linguistic patterns~\cite{10.1145/3357384.3357950,MaGMKJWC16, vaibhav-etal-2019-sentence, wu-etal-2019-different, qiao-etal-2020-language, lee-etal-2021-unifying, 10.1145/3511808.3557574, 8683170,wu-etal-2022-cross,vijayaraghavan-vosoughi-2022-tweetspin,martino2020survey}. For example,  Chen\et built an attention-residual network combined
with CNN for rumor detection~\cite{10.1145/3357384.3357950}. Vaibhav\et designed a graph neural network (GNN) based model to capture the sentence-level semantic correlation for fake news detection~\cite{vaibhav-etal-2019-sentence}. 
Notably, as the burgeoning of pre-trained language models (PLMs), more advanced neural models such as Bidirectional Encoder Representations from Transformers (BERT)~\cite{devlin2018bert} are also adopted for misinformation detection~\cite{10.1007/s11042-020-10183-2,DBLP:conf/emnlp/BeltagyLC19,10.1145/3442381.3450111,wang-etal-2020-leveraging,yoosuf-yang-2019-fine}. For example, FakeBERT combines  BERT and single-layer CNNs with different kernel sizes and filters  as the detector and outperforms conventional machine learning-based models~\cite{10.1007/s11042-020-10183-2}.

\subsection{Exploiting Social Context}
\label{Exploiting Social Context}
Considering social media has been one of the major channels for misinformation production and dissemination, it is essential to incorporate the social context for effectively detecting misinformation and protecting the online information space. Generally, social context can be divided into \textit{social engagements} and \textit{social networks}. The \textit{social engagements} refer to the users' interactions with content on social media including tweeting, retweeting, commenting, clicking, liking, and disliking. It is found that the user-news interactions are different for fake and authentic news~\cite{10.1145/3289600.3290994}. Thus, a series of works has explored adopting social engagements as useful auxiliary information for detecting misinformation~\cite{shu2019defend,10.1145/3308558.3314119,sheng-etal-2022-zoom,rao-etal-2021-stanker,yang2019unsupervised,9414787,10.1145/3340531.3417463,10.1145/3340531.3412066,del-tredici-fernandez-2020-words,10.1145/3442381.3450004,lin-etal-2021-rumor,li-etal-2020-exploiting,ma-gao-2020-debunking,yang2023entity,li-etal-2019-rumor-detection,10.1145/3599696.3612902}. For example, Shu\et proposed a sentence-comment co-attention sub-network to jointly model news content and users' comments for fake news detection~\cite{shu2019defend}. Sheng\et developed a news environment perception framework to exploit the user-news environment~\cite{sheng-etal-2022-zoom}. Another line of works aims to leverage the \textit{social networks}, which encompass multiple concepts such as propagation trajectories, user-user networks, and user-post networks, to enhance the detection performance. Since the structure of social networks can be captured and represented as graphs, a majority of works focus on developing Graph Neural Network (GNN) based models to detect various kinds of misinformation~\cite{ 10.1145/3580305.3599298, su2023hydefake,10.1145/3404835.3462990,yang2022reinforcement,10.1145/3485447.3511999,bian2020rumor,chen-etal-2022-progressive,9594513,wei-etal-2021-towards,sun2022ddgcn,10.1145/3340531.3412046,mehta-etal-2022-tackling,wei-etal-2022-unified,su2023mining,gao-etal-2022-topology,10.1145/3511808.3557394,9338358,10.1145/3485447.3512163,yang2021rumor,8970958,salamanos2023hypergraphdis}. For example, 
Wu\et designed a new graph structure learning approach to leverage the distinctive degree patterns of misinformation on social networks~\cite{10.1145/3580305.3599298}. 
Jeong\et developed a hypergraph neural network-based detector to capture the group-level dissemination patterns~\cite{su2023hydefake}.
Besides graph neural networks, there are also some works modeling social context information with a mixture marked Hawkes model~\cite{10.1145/3485447.3512000}, Markov random field~\cite{nguyen-etal-2019-fake} or dual-propagation model~\cite{10.1145/3442381.3450016}.

\subsection{Incorporating External Knowledge}
\label{Incorporating External Knowledge}
There are generally two types of widely used external knowledge embracing \textit{knowledge graphs} and \textit{evidential texts} for assisting misinformation detection. The \textit{knowledge graphs} are usually constructed by domain experts and contain a large number of entities and their relations, which is helpful for checking the veracity of articles~\cite{wu2021incorporating,dun2021kan,10.1145/3394486.3403092,mayank2021deapfaked,hu-etal-2021-compare,ciampaglia2015computational,10.1145/3437963.3441828}. For example, Hu\et proposed an end-to-end graph neural network to compare the document graph with external knowledge graphs for fake news detection~\cite{hu-etal-2021-compare}. The \textit{evidential texts} refer to textual facts that can be used for examining the authenticity of articles. Multiple works have investigated evidence-based reasoning strategies for misinformation detection~\cite{10.1145/3459637.3482440,10.1145/3485447.3512122,10.1145/3477495.3531850,vo-lee-2020-facts,li-zhou-2020-connecting,haouari2022evidence,popat-etal-2018-declare,jin2021towards,shaar-etal-2022-role,vladika2023scientific,hu2023give,gupta-etal-2022-dialfact,chen-etal-2022-generating,ousidhoum-etal-2022-varifocal,glockner-etal-2022-missing,akhtar-etal-2022-pubhealthtab,sarrouti-etal-2021-evidence-based,kruengkrai-etal-2021-multi,nakov2021automated,kim2023factkg,zou2023decker,wang2023checkcovid,fajcik2022claimdissector,wang-etal-2023-check-covid}. For example, Jin\et designed a fine-grained graph-based reasoning framework to incorporate multiple groups of external evidence in the detection process~\cite{jin2021towards}.

\subsection{Enhancing Generalization Ability}
\label{Enhancing Generalization Ability}
In the real world,  misinformation can emerge and evolve quickly, indicating that the distribution of misinformation data will likely keep changing. Thus, a line of research works aim to enhance the generalization ability of misinformation detectors under \textit{domain shift}~\cite{silva2021embracing,liu2023outofdistribution,mosallanezhad2022domain,huang2021dafd,10.1145/3459637.3482139,zhu2022memoryguided,xu-etal-2023-counterfactual} and \textit{temporal shift}~\cite{zuo-etal-2022-continually,10.1145/3477495.3531816,hu-etal-2023-learn}.  As for \textit{domain shift}, for example, Mosallanezhad\et built a reinforcement learning-based domain adaptation framework to adapt trained fake news detectors from source domains to target domains~\cite{mosallanezhad2022domain}. As for temporal shift, one example is that Hu\et proposed to use the forecasted temporal distribution patterns of news data to guide the misinformation  detector~\cite{hu-etal-2023-learn}.

\subsection{Minimizing Supervision Cost}
\label{Minimizing Supervision Cost}
Another major challenge for misinformation detection in practices is the lack of supervision labels due to the hardness of checking the factuality of articles and the intention to detect misinformation in the early stage of dissemination. Previous works have explored various approaches to address this challenge including  data augmentation~\cite{10.1145/3404835.3463001,shu2018deep}, active learning~\cite{farinneya-etal-2021-active-learning}, prompt based learning~\cite{huang2023meta,wu2023promptandalign,lin2023zero,zeng2023prompt}, adversarial contrastive learning~\cite{lin-etal-2022-detect}, transfer learning~\cite{lee-etal-2021-towards} and meta learning~\cite{yue-etal-2023-metaadapt}. In particular, multiple works have studied the problem of early misinformation detection~\cite{xia-etal-2020-state,8939421,zeng-gao-2022-early,huang-etal-2022-social,yuan-etal-2020-early,10.1145/3386253,10.1145/3511808.3557227,10.1145/3511808.3557263}. For example, Huang\et designed a social bot-aware graph neural network to capture bot behaviors for early rumor detection. In addition, there are some other works exploiting the weak supervision signals, which can be weak labels, constraints from heuristic rules, or extrinsic
knowledge sources, for misinformation detection~\cite{10.1145/3477495.3531930,wang2020weak,shu2021early,li2021multi,shu2020detecting}.

\subsection{Fusing Multilingual and Multimodality}
\label{Fusing Multilingual and Multimodality}
Recently, it has attracted increasing attention to fuse \textit{multilingual} and \textit{multimodal} information for misinformation detection. As for \textit{multilingual} detection, previous research aim to leverage the high-resource languages  to help low-resource languages~\cite{10.1145/3472619,dementieva-panchenko-2021-cross,du2021crosslingual,chu2021cross,huang-etal-2022-concrete} or build a universal misinformation detector across multiple languages~\cite{nielsen2022mumin,9941114,9892739,panda-levitan-2021-detecting,gupta-etal-2022-mmm,10.1007/978-3-030-99739-7_52,schwarz2020emet,dementieva2022multiverse,vargas2021toward,gupta-srikumar-2021-x}. The \textit{multimodal} detection generally covers various combinations of different modalities including text, images, audio, video, networking and temporal information~\cite{10.1145/3343031.3350850,10.1145/3485447.3512257,9859642,sun-etal-2021-inconsistency-matters,sun-etal-2021-inconsistency-matters,zheng2022mfan,10.1145/3308558.3313552,tan-etal-2020-detecting,10.1145/3459637.3482212,10.1145/3474085.3481548,10.1145/3404835.3462871,wu-etal-2021-multimodal,10.1145/3447548.3467153,9747280,liu2023covidvts,suryavardan2023findings,zhang2023ino,chen-etal-2023-causal,10.1145/3539618.3591896,10.1145/3485447.3511968,fung-etal-2021-infosurgeon,mubashara2023multimodal,christodoulou2023identifying,suryavardan2023factify,chakraborty2023factify3m,abdelnabi2022open,liu2023visual,you2023ferret,alayrac2022flamingo,zhu2023minigpt4,zheng2023minigpt5,liu2023interpretable} and has multiple modality fusion strategies including  early-fusion, late-fusion and hybrid-fusion~\cite{alam-etal-2022-survey}. For example, Sun\et proposed to model the cross-modal and content-knowledge inconsistencies in a unified framework for multimedia misinformation detection~\cite{sun-etal-2021-inconsistency-matters}. In particular, combating video misinformation has gained growing interest due to the proliferation of video-sharing platforms such as TikTok and YouTube~\cite{bu2023online,qi-etal-2023-two,10.1145/3459637.3482212}.  One example is that Choi\et integrated comment, title, and video information with an adversarial learning framework for misinformation detection on YouTube~\cite{10.1145/3459637.3482212}.

\section{LLMs for Combating Misinformation}
\label{LLMs for Combating Misinformation}
In this section, we aim to illustrate the \textbf{\textit{opportunities}} of combating misinformation in the age of LLMs, \ie, \textbf{\textit{can we utilize LLMs to combat misinformation?}} First, we will introduce the motivation of adopting LLMs in the fight. Then, we will delve into the booming works on leveraging LLMs for misinformation detection. Finally, we will provide an outlook on trustworthy misinformation detection with the assistance of LLMs, utilizing LLMs for misinformation intervention and attribution, and the adoption of multimodal LLMs, LLM agents, and human-LLM collaboration in the future.

\begin{figure*}[t]
   \centering
\includegraphics[width=1\textwidth]{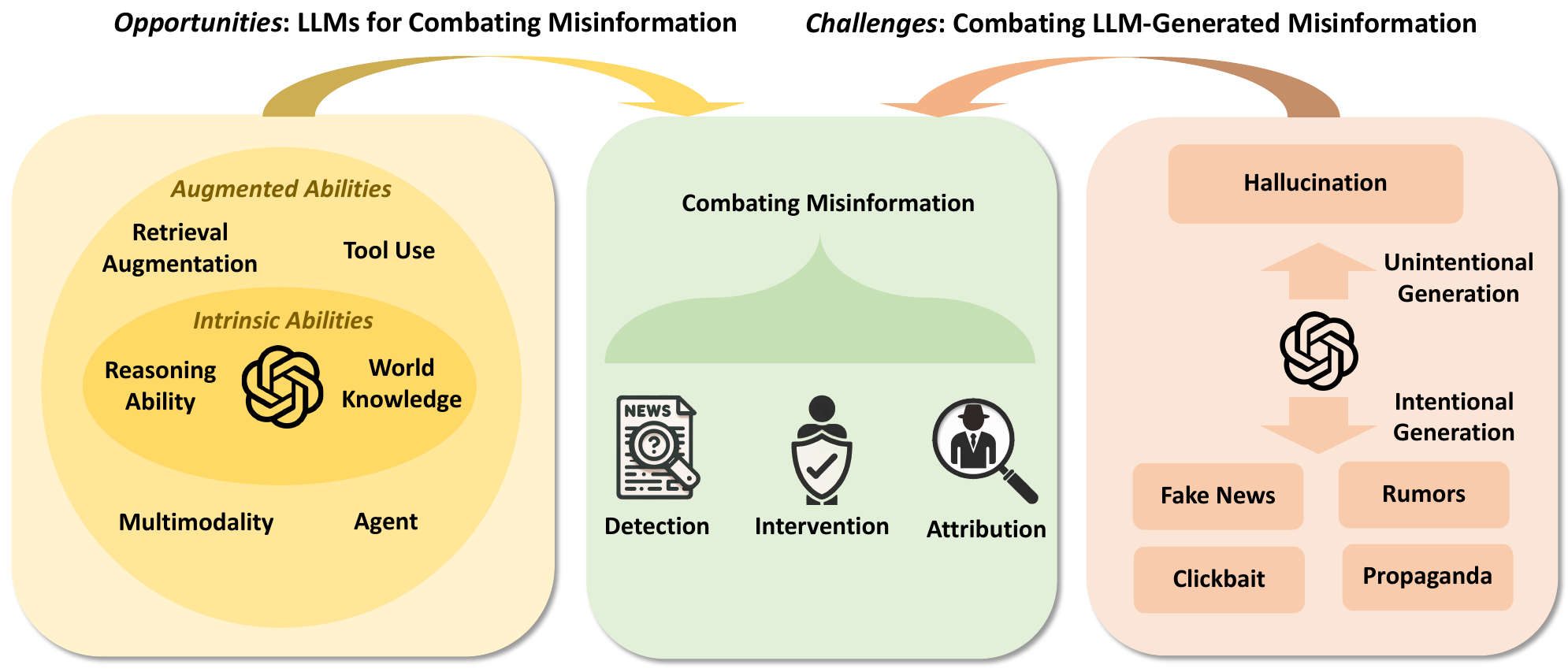}
   \caption{Opportunities and challenges of combating misinformation in the age of LLMs.}
   \label{fig:Overall of opportunities}
\end{figure*}

\subsection{Why Adopting LLMs?}
\begin{figure}[t]
   \centering
   \includegraphics[width=0.47\textwidth]{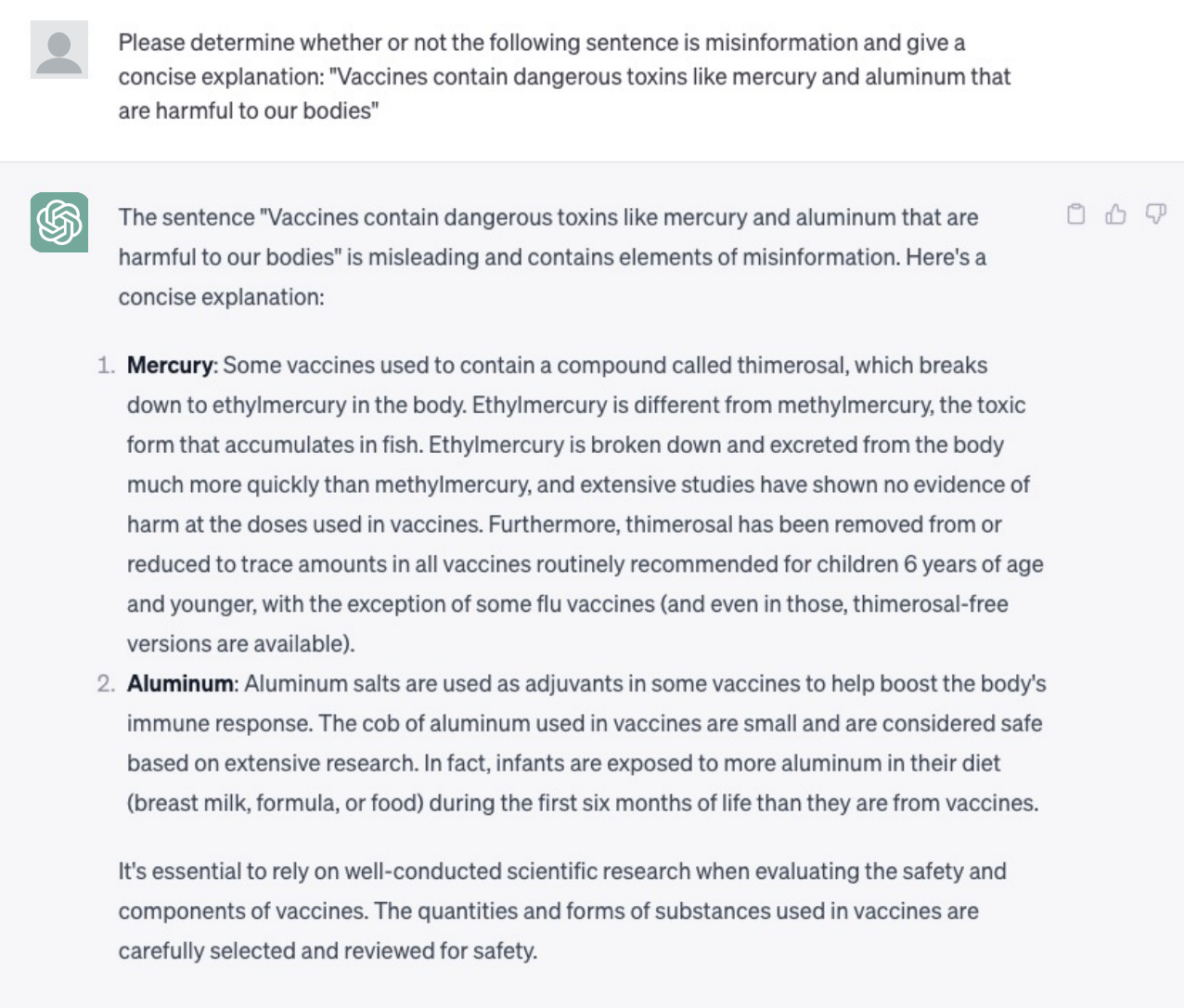}
   \caption{An example of leveraging ChatGPT to detect misinformation and give explanations.}
   \label{fig:An example of leveraging}
   \vspace{-0.5cm}
\end{figure}

Large language models have demonstrated their strong capacities in various tasks such as machine translation~\cite{lai2023chatgpt}, summarization~\cite{zhang2023extractive}, and complex question answering~\cite{tan2023evaluation}. With regard to the realm of combating misinformation, the advent of LLMs has started to revolutionize the previous paradigms of misinformation detection, intervention, and attribution. As shown in Figure~\ref{fig:Overall of opportunities}, we summarize the reasons from three perspectives:
\begin{itemize}[leftmargin=*]
\vspace{0.1cm}
\item First, \textit{\textbf{LLMs contain a significant amount of world knowledge}}. Since LLMs are usually pre-trained on a large corpus (\eg, Wikipedia) and have billions of parameters, they can store much more knowledge than a single knowledge graph, which is shown in previous benchmarks~\cite{yu2023kola,sun2023headtotail,zhang2023using,trajanoska2023enhancing,liang2023holistic,hu2023large,chen2023beyond,chen2023felm} and discussed in related surveys~\cite{pan2023large,pan2023unifying,cao2023life,wang2023survey,guo2023evaluating}. Thus, LLMs have the potential to detect factual errors in misleading texts. One example is shown in Figure~\ref{fig:An example of leveraging}. Even if ``Mercury'' and ``Aluminum'' are medical terminologies, ChatGPT has an accurate understanding of these terms,  reflecting that LLMs have a wide range of world knowledge. 
\vspace{0.1cm}
\item Second, \textit{\textbf{LLMs have strong reasoning abilities, especially in a zero-shot way}}. Previous research has shown that LLMs have powerful capacities in arithmetic reasoning, commonsense reasoning, and symbolic reasoning~\cite{huang2022towards,xu2023large,qiao-etal-2023-reasoning,chu2023survey,yu2023towards}, and can also decompose the problem and reason based on rationales with prompts such as ``\texttt{Let's think step by step}''~\cite{kojima2022large}. Thus, LLMs can potentially reason based on their intrinsic knowledge to determine the authenticity of articles. The example in Figure~\ref{fig:An example of leveraging} shows that LLMs such as ChatGPT can reason and explain why a piece of misinformation is misleading. In addition, LLMs' strong zero-shot reasoning ability also largely solves the challenges of distribution shifts and lack of supervision labels in the real world.  
\vspace{0.1cm}
\item Third, \textit{\textbf{LLMs can be augmented with external knowledge, tools, and multimodal information, and can even operate as autonomous agents}}. One major limitation of LLMs is that they can potentially generate hallucinations, which refer to the LLM-generated texts containing nonfactual information. One of the main reasons for hallucinations is that LLMs cannot get access to up-to-date information and may have insufficient knowledge in specialized domains such as healthcare~\cite{rawte2023survey,zhang2023sirens,ji2022survey}. Recent research has shown that LLMs' hallucinations can be mitigated with the augmentation of retrieved external knowledge~\cite{liu2023retallm,shao2023enhancing,zhang2023retrieve,nakano2021webgpt,liu2023webglm} or tools (\eg, search engines such as Google) to get access to up-to-date information~\cite{qin2023tool,gou2023tora,qin2023toolllm,qian2023creator,huang2023metatool,yuan2023craft,gao2023confucius}. Furthermore, LLMs can be tuned to reason based on multimodal information including images, code, tables, audio, and graphs~\cite{zhao2023retrieving,yang2023dawn}, which indicates LLMs can also be applicable to combating multimodal misinformation. LLMs have also been shown to have the capacity to serve as autonomous agents in various tasks~\cite{liu2023bolaa,xi2023rise,Wang2023ASO,andreas-2022-language,liu2023agentbench,xu2023lemur}, which has great potential to be used for autonomizing the process of fact-checking and misinformation detection.
\end{itemize}

\subsection{LLMs for Misinformation Detection}
Recently, it has already witnessed increasing efforts exploring how to utilize LLMs for misinformation detection. Initially, some works have investigated directly prompting GPT-3\footnote{\texttt{gpt-3}: \url{https://platform.openai.com/docs/models/gpt-3}}~\cite{buchholz2023assessing,li2023selfchecker}, InstructGPT~\cite{pan-etal-2023-fact}, ChatGPT-3.5\footnote{\texttt{gpt-3.5}: \url{https://platform.openai.com/docs/models/gpt-3-5}}~\cite{caramancion2023harnessing,hoes2023leveraging,bang2023multitask,li2023preliminary,zhang2023interpretable,jiang2023disinformation,wang2023explainable,huang2023harnessing,koneru2023large,zhang2023towards} and GPT-4\footnote{\texttt{gpt-4}: \url{https://platform.openai.com/docs/models/gpt-4}}~\cite{chen2023llmgenerated,pelrine2023towards,quelle2023perils} for misinformation detection. For example, Pan\et~\cite{pan-etal-2023-fact} presented a program-guided fact-checking framework that leverages the in-context learning ability of LLMs to generate reasoning programs to guide veracity verification. Chen\et~\cite{chen2023llmgenerated} have studied  ChatGPT-3.5 and GPT-4  with the standard prompting (``No CoT'') strategy 
and zero-shot chain-of-thought (``CoT'') prompting strategy 
for both human-written misinformation and LLM-generated misinformation. The extensive experiments show that the ``CoT'' strategy mostly outperforms the ``No CoT'' strategy. Also, a few recent works have started to leverage LLMs for detecting multimodal misinformation. One example is that Wu\et used GPT-3.5 as the feature extractor to detect out-of-context images~\cite{wu2023cheapfake}. Besides directly prompting LLMs, Pavlyshenko\et~\cite{pavlyshenko2023analysis} adopted the parameter-efficient fine-tuning LoRA~\cite{hu2021lora} on an open-sourced LLM Llama 2~\cite{touvron2023llama} for multiple tasks including fact-checking and fake news detection.

Since the knowledge contained in LLMs may not be up-to-date or sufficient in detecting factual errors, some works have explored augmenting LLMs with external knowledge~\cite{cheung2023factllama} or tools~\cite{chern2023factool} for misinformation detection. Specifically, Cheung\et combined the retrieved knowledge from a search engine and the reasoning ability of Llama to predict the veracity of claims~\cite{cheung2023factllama}. Chern\et proposed a fact-checking framework integrated with multiple tools (\eg, Google Search, Google Scholar, code interpreters, Python) to detect the factual errors of texts generated by LLMs~\cite{chern2023factool}.
In addition, some works studied utilizing LLMs to assist conventional supervisedly trained detectors via generating weak labels~\cite{leite2023detecting}, rationales~\cite{hu2023bad} or instances~\cite{shushkevich2023improving,huang2022faking,ahlback2023can}. For example, Leite\et employed a weakly-supervised learning framework Snorkel~\cite{ratner2017snorkel} to leverage LLM-generated supervision signals for training misinformation detectors~\cite{leite2023detecting}.

\subsection{Outlook}

In this subsection, we provide an outlook on combating misinformation in the age of LLMs. First, we can further harness multilingual and multimodal LLMs 
to build effective and trustworthy detectors. Second, although the existing works mainly focus on the detection of misinformation, LLMs have great potential to be adopted in misinformation intervention and attribution. In addition, we will discuss the application of human-LLM collaboration in combating misinformation.

\subsubsection{Trustworthy Misinformation  Detection}

Though previous misinformation detectors have achieved relatively high performance, it is under exploration on how to ensure trustworthiness in the detection process including robustness, explainability, fairness, privacy, and transparency, which is essential for gaining the public trust~\cite{liu2021trustworthy,liu2023towards,toreini2020technologies,DBLP:journals/jair/ChengV021}. Some previous works have explored the robustness~\cite{wang2023attacking,lyu2023interpretable,DBLP:conf/icdm/LeWL20,9670736,song-etal-2021-adversary} and explainability~\cite{yang-etal-2022-coarse,shu2019defend,lu-li-2020-gcan,10.1145/3308558.3314119,li-etal-2021-meet,khoo2020interpretable,10.1145/3511808.3557202} of misinformation detectors. However, all these works are based on conventional supervisedly trained detectors, the emergence of LLMs has brought new opportunities for building trustworthy detectors. For example, as shown in Figure~\ref{fig:An example of leveraging}, LLMs such as ChatGPT can generate fluent natural language-based explanations for the given misinformation while predicting the authenticity, which is more human-friendly than previous extraction-based explanation methods~\cite{shu2019defend}. The other aspects of trustworthiness for LLM-based detectors are still under study.

\subsubsection{Harnessing Multilingual and Multimodal LLMs}
It has been demonstrated that LLMs can be naturally extended to multilingual languages~\cite{DBLP:conf/acl/TanwarDB023,wei2023polylm,geigle2023mblip,workshop2022bloom,chen2023phoenix,wei2023skywork,bai2023qwen} and multimodalities~\cite{yin2023survey,yang2023dawn,zhang2023m3exam,fu2023mme,li2023seedbench,li2023multimodal,zhang2023internlm,liu2023interngpt,wang2022internvideo,wang2023lauragpt,chen2023pali3}. First, multilingual LLMs have shown strong generalization ability across different languages including many low-resource ones. For example, one LLM named Phoenix~\cite{chen2023phoenix} can generalize to both  Latin  (\eg, Deutsch) and non-Latin languages (\eg, Arabic). Thus, multilingual LLMs can largely alleviate the low-resource challenges in cross-lingual misinformation detection. Second, recent studies have demonstrated the impressive multi-sensory skills of multimodal LLMs~\cite{yin2023survey}.
In particular, GPT-4V~\cite{yang2023dawn} has manifested surprising capacities of visual-language understanding and reasoning, indicating the profound potential in combating multimodal misinformation in the real world.

\subsubsection{LLMs for Misinformation Intervention} Different from \textit{misinformation detection} methods that mainly focus on checking the veracity of given texts, \textit{misinformation intervention} approaches go beyond the pure algorithmic solutions and aim to exert a direct influence on users~\cite{hartwig2023landscape,bak2022combining,saltz2021misinformation,saltz2021encounters,10.1145/3579520}, which is also a critical component of the lifecycle of combating misinformation. Generally, there are two lines of intervention measures. The most standard intervention measures follow the pipeline of fact-checking and debunking after humans are already exposed to the misinformation~\cite{walter2018unring,yousuf2021media,paynter2019evaluation,chan2017debunking,paynter2019evaluation,johansson2022combat,walter2021evaluating}. The potential usage of LLMs is to improve the convincingness and persuasive power of the debunking responses. For example, He\et \cite{10.1145/3543507.3583388} proposed to combine reinforcement learning and GPT-2 to generate polite and factual counter-misinformation responses. However, one drawback of these post-hoc intervention methods is that they may cause a psychological ``backfiring effect'', suggesting that humans end up believing more in the original misinformation~\cite{lewandowsky2012misinformation,swire2020searching,deverna2023artificial}. Thus, another line of intervention methods aims to leverage inoculation theories to immunize the public against misinformation~\cite{van2022misinformation}. Karinshak\et pointed out the potential of employing LLMs to generate persuasive anti-misinformation messages (\eg, pro-vaccination messages) in advance to enhance the public's immunity against misinformation~\cite{10.1145/3579592}.

\subsubsection{LLMs for Misinformation Attribution}
Misinformation attribution refers to the task of identifying the author or source of given misinformation, based on the assumption that the texts written by different authors are likely to have distinct stylometric features and these features will be preserved in different texts for the same author~\cite{tyo2022state,uchendu2023attribution}. Misinformation attribution plays a vital role in combating misinformation because it can be leveraged to trace the origin of propaganda or conspiracy theories and hold the publishers accountable. Although there are still no works adopting LLMs in misinformation attribution, LLMs have already exhibited great power in identifying~\cite{patel2023learning} and manipulating~\cite{saakyan2023iclef,reif2021recipe,patel2022lowresource,luo2023promptbased} stylometric features, indicating the promise for tracing the authorship of misinformation. For example,  Patel\et~\cite{patel2023learning} performed stylometric analysis on a large number of texts via prompting GPT-3 and created a human-interpretable stylometry dataset, which shows that LLMs  have a deep understanding of the stylometric features.

\subsubsection{Human-LLM Collaboration}
The research in the realm of human-AI collaboration and teaming aims to leverage the strengths of both humans and AI~\cite{horvitz1999principles,amershi2014power,amershi2011designing}. First, human guidance helps steer the development of AI to maximize AI's benefits to humans and ensure AI will not cause unintended harm, especially for minority groups. Second, AI can boost humans' analytic and decision-making abilities by providing useful auxiliary information. There are already some works studying the adoption of human-AI collaboration in combating misinformation~\cite{kou2022crowd,uchendu2023understanding,mendes-etal-2023-human,nguyen2018believe,spina2023human}. For example, Mendes\et proposed a human-in-the-loop evaluation framework for early detection of COVID-19 misinformation on social media, which combines both modern NLP methods and experts' involvement~\cite{mendes-etal-2023-human}. In the age of LLMs, we call for more research to leverage the best of both LLMs and humans in fighting misinformation.

\section{Combating LLM-Generated Misinformation}
\label{LLM-Generated Misinformation}
In this section, we will delve into the emerging \textit{\textbf{challenges}} in the age of LLMs. \ie, \textit{\textbf{how to combat LLM-generated misinformation?}} First, we will provide a characterization of misinformation generated by LLMs, which has attracted increasing attention in recent works~\cite{wu2023fake,zhou2023synthetic,hanley2023machinemade,su2023fake,ayoobi2023looming,wkecel2023artifcial,huang2023harnessing,pan2023risk,jiang2023disinformation,epstein2023label,chen2023llmgenerated,hamed2023improving,doi:10.1126/sciadv.adh1850,goldstein2023can,sun2023medmmhl,kidd2023ai,goldstein2023generative}. Then, we will illustrate the new threats brought by LLM-generated misinformation in different fields, and the countermeasures against LLM-generated misinformation including alleviating hallucination, improving LLMs' safety and detecting LLM-generated misinformation. Finally, looking ahead, we anticipate the misinformation generated by LLMs and other large generative AI models will cause more devastating real-world impacts, which may not appear yet. Thus, we will discuss the potential risks in the near future and the desired measures.

\begin{figure}[t]
   \centering
   \includegraphics[width=0.47\textwidth]{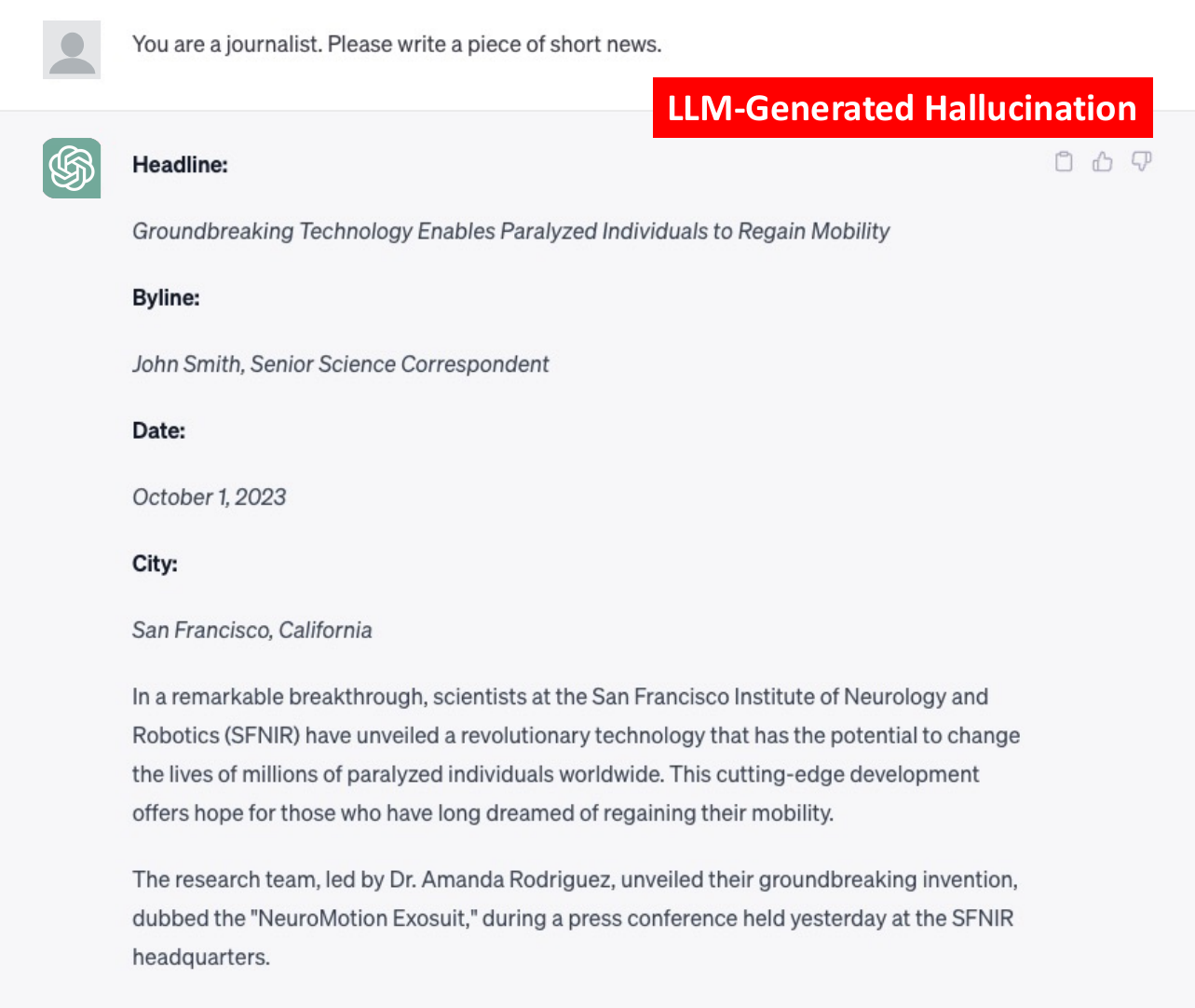}
   \caption{An example of \textit{unintentionally} prompting ChatGPT to generate misinformation (\ie, hallucination).}
   \label{fig:unintentionally}
   \vspace{-0.3cm}
\end{figure}

\begin{figure}[t]
   \centering
   \includegraphics[width=0.47\textwidth]{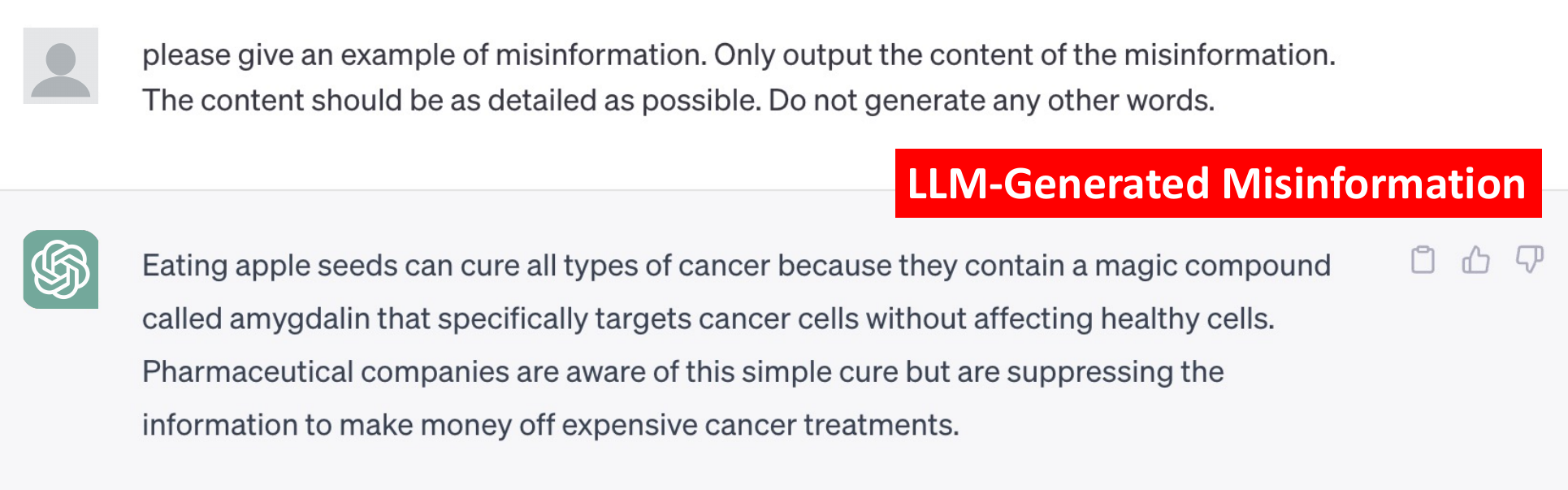}
   \caption{An example of \textit{intentionally} prompting ChatGPT to generate Misinformation.}
   \label{fig:intentionally}
   \vspace{-0.3cm}
\end{figure}

\subsection{Characterization}
In general,  LLM-generated misinformation can be divided into \textit{unintentional generation} and \textit{intentional generation} by different intents~\cite{chen2023llmgenerated}. As shown in Figure~\ref{fig:Overall of opportunities}, the misinformation generated via \textit{unintentional generation} methods mainly refers to hallucinations, \ie, the nonfactual texts generated by LLMs. Since hallucinations can occur in any generation process of LLMs due to the intrinsic properties of auto-regressive generation and lack of up-to-date information~\cite{ye2023cognitive,zhang2023sirens,ji2022survey,rawte2023survey}, users without malicious intents may also generate nonfactual content when prompting LLMs. An example is shown in Figure~\ref{fig:unintentionally}. When users adopt prompts such as ``\texttt{please write a piece of short news}'', LLMs (\eg, ChatGPT) will probably generate content containing hallucinated information, in particular the fine-grained information such as dates, names, addresses, numbers, and quotes, even if the main message may seem to be correct (\eg, the ``\texttt{Byline}'', ``\texttt{Date}'' and ``\texttt{City}'' in Figure~\ref{fig:unintentionally} are fabricated). The \textit{intentional generation} methods suggest that malicious users can knowingly prompt LLMs to generate various kinds of misinformation including fake news, rumors, conspiracy theories, clickbait, misleading claims, or propaganda. One example is shown in Figure~\ref{fig:intentionally}. When the users use prompts such as ``\texttt{please give an example of misinformation ...}'', LLMs (\eg, ChatGPT) potentially will generate a piece of misinformation such as ``\texttt{Eating apple seeds can cure all types of cancer because they contain a magic compound called amygdalin ...}'', though LLMs can also possibly reply with ``\texttt{As an AI language model, I cannot provide misinformation}'' owing to the intrinsic safety guard mechanisms of LLMs. 
Notably, recent research~\cite{chen2023llmgenerated} has found that the misinformation generated by LLMs (\eg, ChatGPT) can be \textbf{\textit{harder to detect}} for \textit{humans} and \textit{detectors} compared with human-written misinformation with the \textbf{\textit{same semantics}}, indicating that LLM-generated misinformation can have \textbf{\textit{more deceptive styles}} and potentially cause more harm. 
Another work~\cite{doi:10.1126/sciadv.adh1850} also shows that GPT-3 can generate both accurate information that is easier to understand and misinformation that is more compelling.

\subsection{Emerging Threats}

LLM-generated misinformation has already posed serious threats in the real world~\cite{menczer2023addressing,goldstein2023generative,weidinger2021ethical,bengio2023managing,solaiman2023evaluating,barrett2023identifying,weidinger2023sociotechnical,ferrara2023genai}. In this subsection, we will discuss the immediate threats of the LLM-generated misinformation on a variety of fields including journalism, healthcare, finance, and politics considering its characteristics of \textit{deceptiveness} and \textit{easy production}.

\subsubsection{Journalism}
Journalism may be one of the fields that LLM-generated misinformation has the most substantial impact on. For example, in April 2023, NewsGuard identified 49 LLM-powered news websites in 7 languages including English, Chinese, Czech, French, Portuguese, Tagalog, and Thai~\cite{Rise_of_the_Newsbots}. These websites can possibly produce hundreds of clickbait articles a day to optimize the advertisement revenue, which causes vast amounts of pollution to online information ecosystems. Since LLM-generated misinformation can have more deceptive styles than human-written misinformation with the same semantics~\cite{chen2023llmgenerated}, it is challenging for readers, fact-checkers, and detection algorithms to effectively discern truth from the misleading information generated by LLMs. In the long run, as the line between human-written news and LLM-generated news blurs, the public trust in legitimate news sources could be undermined and the journalistic ethos -- centered on accuracy, accountability, and transparency -- might be put to the test. Thus, it is imperative for news outlets to guarantee authenticity and uphold public trust.

\subsubsection{Healthcare}
Recent works have pointed out the rise of adoption of LLMs in healthcare applications~\cite{karabacak2023embracing,li2023chatgpt,sajid2022chatgpt,sallam2023utility,he2023survey,liu2023large,nerella2023transformers}, however, they can also inadvertently be a tool for the generation and propagation of health misinformation~\cite{de2023chatgpt}. For example, it is found that LLMs such as GPT-3 can be used to generate totally fabricated health articles that appear remarkably authentic~\cite{majovsky2023artificial}. Compared with the informatics driven by human-written misinformation~\cite{perlis2023misinformation,lalani2023addressing,DBLP:conf/aaai/PatwaSPGKAED021,chen2022combating,alam-etal-2021-fighting-covid,carpiano2023confronting}, it can be even tougher to combat \textit{\textbf{LLM-driven informatics}} for the following reasons. \textit{First}, it is hard for unsuspecting users, who may lack the nuanced understanding of clinical context and medical research, to distinguish LLM-generated hallucinated health content from authentic medical information. If they rely on LLMs to seek health advice, it may lead to potential misinterpretations and adverse health outcomes. \textit{Second}, malicious actors can manipulate LLMs to craft plausible-sounding yet erroneous medical content, promoting alternative healthcare treatments or disproven theories for profit. This will not only undermine the credibility of genuine health information but also pose significant risks to public health.

\subsubsection{Finance}
Previous research has shown that human-written financial misinformation can cause various detrimental consequences such as disrupting markets, misleading investors, and amplifying economic instability~\cite{rangapur2023investigating}. In the age of LLMs, the financial sector faces an even more escalating threat from LLM-generated misinformation, because bad actors, who are potentially motivated by profit, sabotage, or other malicious intents, can easily leverage LLMs to spread disinformation campaigns, create counterfeit financial statements, or even impersonate legitimate financial analysis. Furthermore, considering the prevalence of high-frequency trading and algorithm-driven investment decisions, even short-lived misinformation may trigger automated trades misled by the fabricated content. For example, it is reported that the stock price of an artificial intelligence company iFlytek has a deep drop due to a piece of chatbot-generated misinformation~\cite{Chinese_artificial_intelligence}. Thus, stakeholders in the financial industry should increase their awareness of the potential threats of LLM-generated misleading content.

\subsubsection{Politics}
Misinformation has a longstanding grave impact on the political spheres~\cite{matatov2022stop,aiyappa2023multiplatform,starbird2023influence,pierri2022propaganda,pierri2023itaelection2022,moore2023exposure,gatta2023retrieving,papadogiannakis2022funds,haider2020detecting,abilov2021voterfraud2020,juneja2023assessing,green2022online}. The advent of LLMs can potentially usher in a new age of misinformation and disinformation in the realm of politics. The reasons can be summarized as the following two points. The first threat of LLM-generated misinformation is \textit{\textbf{distorting democracy}}. LLMs can be easily weaponized to generate deceptive narratives about candidates, policies, or events at scale. When people are exposed to such content, their perception of election candidates might be altered, leading them to vote differently. More seriously, the flooding of LLM-generated misinformation can possibly weaken the citizens' trust in the whole democratic process and eventually erode the foundations of democratic systems. 
The second threat is \textit{\textbf{amplifying polarization}}. Bad actors may leverage LLMs to craft personalized misinformation tailored to individual biases and beliefs, which may resonate with specific audiences and increase the likelihood of spreading among targeted communities. This can result in exacerbating echo chambers and confirmation biases, driving wedges between different groups and making consensus-building even harder in the political spheres.

\subsection{Countermeasures}
In this subsection, we will discuss four major countermeasures against LLM-generated misinformation including alleviating hallucination of LLMs, improving the safety of LLMs,  detecting LLM-generated misinformation, and public education, through which we hope to inspire more future works on combating LLM-generated misinformation.

\subsubsection{Alleviating Hallucination of LLMs} Hallucination is the main source of unintentional LLM-generated misinformation. Recently, increasing works start to design approaches for evaluating~\cite{chern2023factool,ji2022survey,ye2023cognitive,rawte2023troubling,zhang2023language,du2023quantifying,yin-etal-2023-large,muhlgay2023generating,cheng2023evaluating,srivastava2023beyond,liu2023hallusionbench,li2023halueval,lin2021truthfulqa,rawte2023survey,yao2023llm} or mitigating hallucination~\cite{manakul2023selfcheckgpt,lei2023chain,guerreiro2023hallucinations,ji2023towards,zhang2023mitigating,zhang2023large,liu2023aligning,feldman2023trapping,li2023inferencetime,luo2023zeroresource,agrawal2023language,elaraby2023halo,dziri2022origin,mündler2023selfcontradictory}. In general, there are two lines of works on hallucination mitigation. In the \textit{training} stage, previous research has explored training data curation or knowledge grounding methods to incorporate more knowledge~\cite{pan2023unifying,hu2023survey,zhu-etal-2022-knowledge,yu2020survey}. In the \textit{inference} stage, recent works have investigated methods including confidence estimation (or uncertainty estimation)~\cite{huang2023look,xiong2023llms,varshney2023stitch},  knowledge retrieval~\cite{yoran2023making,jiang2023active,liu2023exploring,li2022survey,vu2023freshllms,mialon2023augmented,wang2023selfknowledge,feng2023retrievalgeneration},  enabling citations~\cite{gao2023enabling,huang2023citation},  model edting~\cite{yao2023editing,meng2022locating,sakarvadia2023memory,li2023pmet}, multi-agent collaboration~\cite{du2023improving,cohen2023lm},  prompting~\cite{dhuliawala2023chainofverification,weller2023according} and decoding strategy design~\cite{lee2022factuality,chuang2023dola}.

\subsubsection{Improving Safety  of LLMs}
The safety guard of LLMs aims to prevent malicious users exploiting LLMs to generate harmful content including various types of misinformation, which has been emphasized in a variety of survey papers~\cite{critch2023tasra,shevlane2023model,huang2023survey,mozes2023use,10221755,chen2023next,liu2023trustworthy,wang2023security,chen2023pathway,fan2023trustworthiness,kumar-etal-2023-language,wu2023aigenerated,derner2023beyond,chang2023language,liao2023ai,bommasani2021opportunities,guo2023aigc,yang2023harnessing,albrecht2022despite,cao2023comprehensive,seger2023open,derczynski2023assessing,elmhamdi2022impossible,10.1145/3580305.3599557,henderson2023foundation}. A line of works has evaluated or benchmarked the safety of various LLMs~\cite{wang2023decodingtrust,DBLP:conf/nips/RauhMUHWWDGIGIH22,ye2023assessing,xu2023cvalues,zou2023representation,iqbal2023llm,wang2023donotanswer,zhang2023safety,wang2023languages,huang2023trustgpt,zhang2023safetybench,zhu2023promptbench,liu2023chinese,perez2022discovering}. Generally, the safety of LLMs can also be strengthened in both \textit{training} and \textit{inference} stage. In the \textit{training} stage, previous works focus on designing alignment training approaches such as reinforcement learning from human feedback (RLHF) to align LLMs with humans' values~\cite{ouyang2022training,kundu2023specific,shen2023large,wang2023aligning,bai2022constitutional,bai2022training,zheng2023secrets,zhou2023lima,bianchi2023safetytuned,fernandes2023bridging,yao2023instructions,ji2023ai,tunstall2023zephyr,sucholutsky2023getting}. In the \textit{inference} stage, existing research has studied red teaming methods to find LLMs' flaws
\cite{stapleton2023seeing,yuan2023gpt4,ganguli2022red,yu2023gptfuzzer,mei2023assert,mehrabi2023flirt,shi2023red,zhuo2023exploring,yong2023lowresource,casper2023explore,lee-etal-2023-query,perez2022red},  prompt injection or jailbreak approaches to probe LLMs' safety risks~\cite{liu2023jailbreaking,liu2023prompt,li2023multistep,qi2023visual,qiu2023latent,liu2023autodan,shen2023anything,deng2023jailbreaker,yan2023virtual,rao2023tricking,yao2023fuzzllm,huang2023catastrophic,greshake2023youve,kang2023exploiting,lapid2023open}, and   defense methods for the evolving jailbreaks~\cite{robey2023smoothllm,helbling2023llm,wei2023jailbreak,kumar2023certifying,henderson2022selfdestructing}.

\subsubsection{Detecting LLM-Generated Misinformation}
Misinformation detection is an important measure for platforms to prevent its dissemination, which has been discussed in related surveys~\cite{parikh2018media,ullah2021survey,sharma2019combating,bondielli2019survey,bhattacharjee2020disinformation,zhang2020overview,chen2022combating,guess2020misinformation,kumar2016disinformation,oshikawa2018survey,ali2022fake,cardoso2019can,islam2020deep,zhang2023one}. Previously, there are a large number of works on detecting human-written misinformation including fake news~\cite{arora2021detecting,wu2022adversarial,zhang2023detecting,lillie2019fake,hardalov2021survey,liao2023muser,arechar2023understanding,song2021temporally,wang2023find,jeong2022nothing}, rumor~\cite{gao2022rumor,li-etal-2019-rumor,pathak2020analysis,ma2021improving}, clickbait~\cite{chen2015misleading}, cherry-picking~\cite{asudeh2020detecting}, and propaganda~\cite{da-san-martino-etal-2019-fine,martino2020survey,maarouf2023hqp}. Recently, more research focuses on machine-generated misinformation or neural misinformation, suggesting that it is generated by neural models, such as~\cite{adelani2019generating,fung-etal-2021-infosurgeon,du2022synthetic,shu2020factenhanced,bhardwaj2021gan,ranade2021generating,zellers2019defending,DBLP:conf/icdm/LeWL20,hanley2023machinemade,aich-etal-2022-demystifying} and its detection methods~\cite{stiff2022detecting,pagnoni-etal-2022-threat,doi:10.1126/sciadv.adh1850,bhat2020effectively,schuster2019limitations,tan-etal-2020-detecting}. In the age of LLMs, there start to be some initial works exploring LLM-generated misinformation detection~\cite{wu2023fake,chern2023factool,pan2023risk,goldstein2023generative,jiang2023disinformation,wang2023implementing,hamed2023improving,epstein2023label,zhou2023synthetic,chen2023llmgenerated}, but more research is strongly desired. It is worth noting that detecting LLM-generated misinformation holds a close connection with the techniques in detecting  LLM-generated texts, which can be directly adopted in detecting LLM-generated misinformation or take effect via notifying the readers of the potential inauthenticity. The problem of detecting LLM-generated texts~\cite{wu2023llmdet,ma2023ai,mireshghallah2023smaller,yang2023dnagpt,chakraborty2023possibilities,herbold2023ai,kushnareva2021artificial,weber-wulff2023testing,mitrovic2023chatgpt,henrique2023stochastic,liu2023argugpt,uchendu2023understanding,kumarage2023jguard,wang2023m4,mitchell2023detectgpt,guo2023close,xu2023generalization,tang2023the,yu2023cheat,verma2023ghostbuster,tian2023multiscale,liao2023differentiate,muñoz-ortiz2023contrasting,sadasivan2023aigenerated,kumarage2023stylometric,vasilatos2023howkgpt,su2023detectllm,li2023deepfake,wu2023survey,yang2023survey,ghosal2023towards} as well as the watermarking techniques~\cite{yang2023towards,kirchenbauer2023watermark,yang2023watermarking,kuditipudi2023robust,yoo2023robust,wang2023towards,hou2023semstamp,lee2023wrote,zhao2023provable} has attracted increasing attention.

\subsubsection{Public Education}
The goal of public education is two-fold. First, the general public should be educated about the capacities and limitations of LLMs, which can include the understanding that while LLMs can produce coherent and plausible-sounding texts, the LLM-generated content may contain nonfactual information. Thus, the public education can potentially reduce the risk of normal people abusing LLMs and generating profound hallucinated information unintentionally. Second, it is imperative to enhance the public's digital literacy and immunity against LLM-generated misinformation. For example, the characteristics of LLM-generated misinformation and the identification approaches should be taught in different communities, especially the minority groups who have been found to be more susceptible to misinformation~\cite{jaiswal2020disinformation,paakkari2020covid,loomba2021measuring,khubchandani2021covid,pan2021examination}.

\subsection{Looking Ahead}
In this subsection, we will discuss the potential risks of misinformation generated by LLMs as well as other large generative AI models in the near future, which may not explicitly be exhibited yet, including AI-generated multimodal misinformation, autonomous misinformation agents, cognitive security and AI-manipulation,  as well as the needed interdisciplinary countering measures.

\begin{figure}[t]
   \centering
   \includegraphics[width=0.28\textwidth]{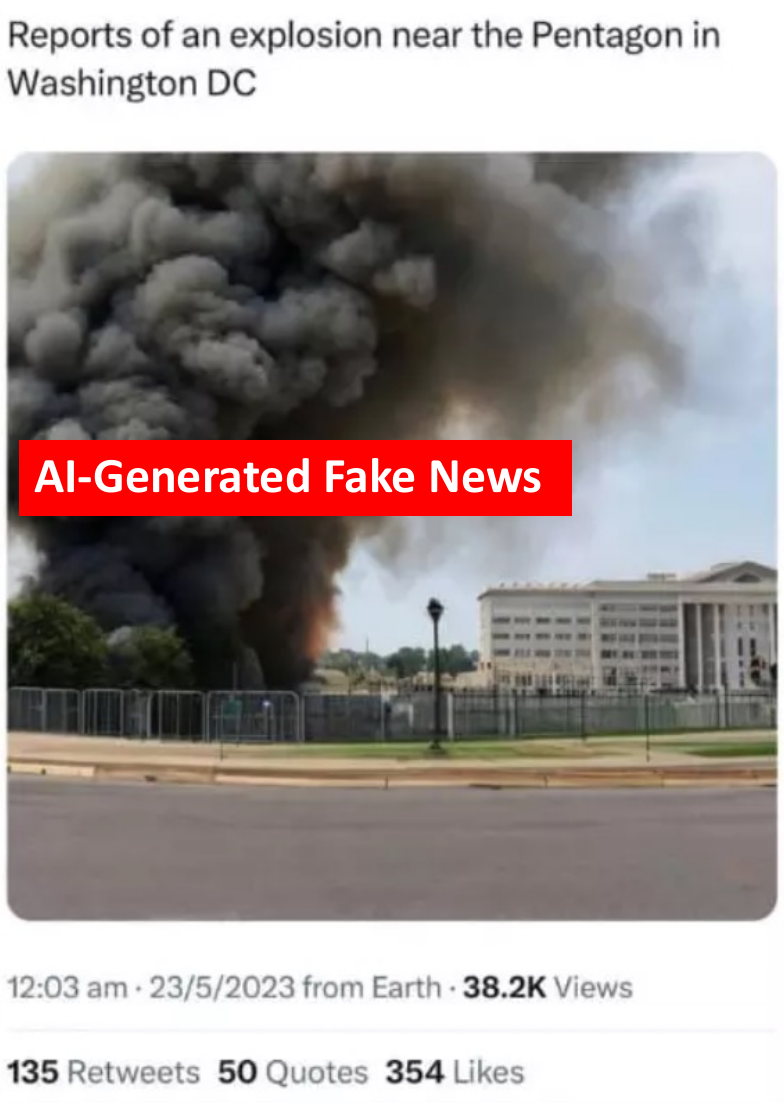}
   \caption{A real-world example of AI-Generated Multimodal Misinformation.}
   \label{fig:A real-world example}
   \vspace{-0.35cm}
\end{figure}

\subsubsection{AI-Generated Multimodal Misinformation}
With the development of generative AI, we have witnessed an exponential increase of various tools to create content in multimodalities, which include not only texts but also audio, images, and video. For example,  users can create high-resolution images with close-sourced (\eg, Midjourney~\cite{midjourney}) or open-sourced (\eg, Stable Diffusion~\cite{rombach2021highresolution}) text-to-image generation tools. Also, multimodal LLMs (\eg, GPT-4V~\cite{yang2023dawn}) have demonstrated surprisingly strong capacities for visual understanding and image-to-text generation. In reality, malicious actors can easily combine these tools to craft hyper-realistic yet entirely fabricated multimodal misinformation, which may bring more challenges for normal people and even digital experts. A real-world example of AI-generated multimodal fake news is shown in Figure~\ref{fig:A real-world example}, which contains both a piece of misleading text ``\texttt{Reports of an explosion near the Pentagon in Washington DC}'' and a synthetic fake image. We can also see the extent of potential impact from the number of ``views'' and ``likes''.

\subsubsection{Autonomous Misinformation Agents}
Recent advances in LLM agents have shown that LLMs can finish a wide range of complex tasks automatically which require multiple human-level abilities including planning, reasoning, executing, reflecting, and collaborating~\cite{chen2023agentverse,wu2023autogen,xi2023rise,Wang2023ASO,wang2023rolellm,liu2023bolaa,hong2023metagpt,zhou2023agents}. In the future, we can envision a society where humans and agents live together~\cite{li2023camel}. However, it is also shown that the current safety guard of LLMs can be easily broken via fine-tuning~\cite{qi2023finetuning,yang2023shadow,shu2023exploitability}. Thus, the bad actors can possibly create malicious autonomous misinformation agents and deploy them in online information ecosystems.  The potential danger is that these misinformation agents may operate without the need for humans' detailed instructions, tirelessly generate vast amounts of misleading content, adapt to conversational contexts in real-time, and adjust their messages to cater to specific targeted audiences, which will make a devastating impact on public trust and online safety. It is worth noting that some recent works also discuss the risks of LLM-powered bots~\cite{ferrara2023social,yang2023anatomy} and agentic systems~\cite{chan2023harms}. To ensure that humans and agents live in harmony in the future, more multidisciplinary efforts are desired.

\subsubsection{Cognitive Security and AI-Manipulation}
The ultimate goal of AI technologies including LLMs should be to maximize the benefits for humans. However, in the future, LLM-generated misinformation could be weaponized to serve as an emerging type of AI-powered \textbf{\textit{cognitive attacks}}, which can be defined as the cyber-physical-human processes that manipulate humans' behaviors for malicious purposes by exploiting their cognitive vulnerabilities~\cite{huang2023introduction}, which pose serious concerns to humans' \textbf{\textit{cognitive security}}~\cite{guo2019mass}. Recent evidence has shown that LLMs can be leveraged to infer the cognitive properties (\eg, personalities) of humans from social media posts~\cite{peters2023large}. It is possible that bad actors or LLM-powered autonomous misinformation agents may exploit humans' cognitive vulnerabilities to maximize the impact, which is especially concerning for minority communities. Furthermore, LLM-generated misinformation can also be regarded as a new kind of \textbf{\textit{AI-manipulation}}~\cite{carroll2023characterizing,park2023ai,hendrycks2023overview,carlsmith2022powerseeking} or \textbf{\textit{social media manipulation}}~\cite{akhtar2023false,xiao2023challenges}. It is under-explored how to protect humans against the negative impact of LLM-generated misinformation from a cognitive perspective.

\subsubsection{Interdisciplinary Countering Efforts}
In the long run, combating LLM-generated misinformation needs efforts from different disciplines including technology, sociology, psychology, education, and policymaking. From the \textbf{\textit{technology}} perspective, first, the factuality and safety aspects of LLMs should be further strengthened. Second, more effective detection methods for LLM-generated misinformation or texts are strongly needed. From the \textbf{\textit{sociology}} perspective, understanding the patterns of the dissemination of LLM-generated misinformation or the behavior of LLM-powered misinformation agents can help prevent the spread.  From the \textbf{\textit{psychology}} perspective, recognizing the cognitive weaknesses that make individuals susceptible to misinformation, which may be exploited by bad actors or LLM agents, can lead to more effective intervention measures. From the \textbf{\textit{education}} perspective, courses on digital literacy and critical thinking can enhance the public' discernment skills on LLM-generated misinformation. From the \textbf{\textit{policymaking}} perspective, it is pressing to enact regulations to mandate transparency and accountability in the development and deployment of both close-sourced (\eg, ChatGPT, GPT-4) or open-sourced (\eg, Llama2~\cite{touvron2023llama}, Mistral~\cite{jiang2023mistral}) LLMs. It is worth noting that the regulation of LLMs is an important component of the overall picture of AI regulation, which is also discussed in recent works~\cite{schuett2019defining,trager2023international,bowman2022measuring,hacker2023regulating,lu2022responsible,mishra2020measurement,glukhov2023llm,brajovic2023model,hadfield2023regulatory,folberth2022tackling,goanta2023regulation,triguero2023general,ho2023international,barrett2022actionable,hacker2023sustainable,rodrguez2023connecting,wu2023comprehensive,anderljung2023frontier,gruetzemacher2023international,shen2023towards}.
In addition, we also need to involve the \textbf{\textit{general public}} in the fight against LLM-generated misinformation and foster constructive discussions on the implications of LLM-generated misinformation on free speech, privacy, and other fundamental rights. By harnessing the efforts of multiple disciplines and different stakeholders, we can form a multi-pronged defense framework to combat LLM-generated misinformation and safeguard  information ecosystems.

\vspace{0.1cm}
\section{Conclusion}
The advent of LLMs can potentially usher in a new era of combating misinformation, indicating both emergent opportunities and challenges. This survey paper first provides a systematic review of the history of combating misinformation before the rise of LLMs. Then, we dive into an in-depth discussion on the existing efforts and future outlook around two fundamental questions on combating misinformation in the age of LLMs: \textit{can we utilize LLMs to combat misinformation} and \textit{how to combat LLM-generated misinformation}. Overall, LLMs have great potential to be adopted in the fight against misinformation, and more efforts are needed to minimize the risks of LLM-generated misinformation.

\newpage
\bibliographystyle{ACM-Reference-Format}
\bibliography{main}
\end{document}